\documentclass[11pt]{article}      
\usepackage{graphicx}
\usepackage{epstopdf}
\begin{document}

\author{Preston Jones \\
\textit{Department of Physics, California State University}\\
\textit{Fresno, California 93740}\\
\textit{electronic mail pjones@csufresno.edu.\bigskip } \and Lucas F. Wanex 
\\
\textit{Department of Physics, University of Nevada}\\
\textit{Reno, Nevada 89557}\\
\textit{electronic mail lucas@physics.unr.edu.}\\}
\title{The Clock Paradox in a Static Homogeneous Gravitational Field}
\maketitle

\begin{abstract}
The \textit{gedanken} experiment of the clock paradox is solved exactly
using the general relativistic equations for a static homogeneous
gravitational field. We demonstrate that the general and special
relativistic clock paradox solutions are identical and in particular that
they are identical for \textit{finite} acceleration. Practical expressions
are obtained for proper time and coordinate time by using the destination
distance as the key observable parameter. This solution provides a formal
demonstration of the identity between the special and general relativistic
clock paradox with finite acceleration and where proper time is assumed to
be the same in both formalisms. By solving the equations of motion for a
freely falling clock in a static homogeneous field elapsed times are
calculated for realistic journeys to the stars.\bigskip

\end{abstract}

\section{Introduction}

Many mathematical solutions to the clock paradox are based on the formalism
of special relativity. This solution parameterizes hyperbolic motion in
terms of the proper time of an accelerating clock\cite{Misner et al}. For
decades the solution to the twin paradox, in terms of the formalism of
general relativity, was also parameterized in terms of this proper time.
Using this parameterization the general relativistic solution has been shown
to be consistent with the special relativistic solution in the limit of
infinite acceleration\cite{Moller}. Recently the clock paradox has been
solved in the formalism of general relativity for finite accelerations with
the maximum relative velocity\cite{Iorio} as the parameter in the solution.
Here we demonstrate that the general and special relativistic clock paradox
solutions are identical with \textit{finite} as well as infinite
acceleration by using the destination distance as the key observable
parameter.

Having reached its one hundredth anniversary the theory of relativity has
become a cornerstone of modern physics. One of the earliest challenges to
this theory was the clock paradox or twin paradox. While this paradox has
long been resolved it remains an excellent vehicle for the study of
relativity and in particular of the relationship between special and general
relativity. The clock paradox is described by a \textit{gedanken} experiment
using identical twins with the first twin remaining at home and the second
twin accelerating away and later returning. In order to describe a
physically realistic trip the accelerating twin's journey will consist of
four legs. During the first leg the twin travels in a space vehicle at a
constant acceleration towards a distant star. The second leg begins where
the traveling twin reverses thrust with constant deceleration. This
deceleration causes the space ship to eventually come to rest relative to
the distant star and the stay at home twin. The third leg consists of
constant acceleration back toward home. Finally the fourth leg begins with
constant deceleration and continues until the traveling twin is reunited
with the stay at home twin.

The principle of equivalence allows this journey to be examined from a
different point of view\cite{Einstein}. From the perspective of the
accelerating twin the stay at home twin is moving in a static and
homogeneous gravitational field (SHF). The equations of motion for a freely
falling clock in this SHF, can be solved in terms of the coordinate distance
of the accelerated twin for both leg 1 and 2 of the out going trip. With the
destination distance in the SHF as a parameter, the two solutions can then
be connected to obtain the elapsed time for a complete trip. Using
coordinate distance as the parameter a coordinate transformation to a
Lorentz inertial frame makes it possible to evaluate elapsed times and
distances for realistic trips to the stars.

\section{Static Homogeneous Field}

Following Einstein's development of the equivalence principle\cite{Einstein}%
\cite{Einstein07} the acceleration experienced by the traveling twin is
indistinguishable from station keeping (remaining motionless) in a
gravitational field. From this perspective the traveling twin is station
keeping while the stay at home twin falls freely in the field. Consider an
orthogonal coordinate system with the gravitational field parallel to the $z$%
-axis. During the first and forth legs of the trip the field points in the
negative direction. During the second and third legs the field is reversed
and points in the positive direction.

The metric of a homogeneous field will be static and unchanged by coordinate
transformations in any plane perpendicular to the acceleration. Under these
conditions all derivatives in the field equations are zero except those
parallel to the acceleration ($z$ axis). This allows the metric for a SHF to
be written in the form,

\begin{equation}
-ds^{2}=c^{2}d\tau ^{2}=V\left( z\right)
^{2}c^{2}dt^{2}-dx^{2}-dy^{2}-U\left( z\right) ^{2}dz^{2}.  \label{1}
\end{equation}

\noindent where $\tau $ is the proper time (time on a freely falling clock), 
$t$ and $x,\ y,\ z$ are the coordinate time and spatial coordinates of the
station keeping twin.

The relation between $V\left( z\right) $ and $U\left( z\right) $ can be
obtained from Einstein's field equations

\begin{equation}
R_{\mu \nu }=-\frac{8\pi G}{c^{4}}\left( T_{\mu \nu }-\frac{1}{2}g_{\mu \nu
}T_{\lambda }^{\lambda }\right) .  \label{2}
\end{equation}

\noindent where $R_{\mu \nu }=0$ in a source free region. The components of
the Ricci tensor are,

\begin{equation}
R_{\mu \nu }=-\frac{\partial }{\partial x^{\alpha }}\Gamma _{\mu \nu
}^{\alpha }+\Gamma _{\mu \beta }^{\alpha }\Gamma _{\nu \alpha }^{\beta }+%
\frac{\partial }{\partial x^{\nu }}\Gamma _{\mu \alpha }^{\alpha }-\Gamma
_{\mu \nu }^{\alpha }\Gamma _{\alpha \beta }^{\beta },  \label{3}
\end{equation}

\noindent and the Christoffel symbols in this equation are\cite{Weinberg},

\begin{equation}
\Gamma _{\mu \nu }^{\sigma }=\frac{1}{2}g^{\sigma \alpha }\left( \frac{
\partial g_{\mu \alpha }}{\partial x^{\nu }}+\frac{\partial g_{\nu \alpha }}{
\partial x^{\mu }}-\frac{\partial g_{\mu \nu }}{\partial x^{\alpha }}\right)
.  \label{4}
\end{equation}

\noindent Using the requirement that only derivatives with respect to $z$
are non-zero leads to,

\begin{equation}
\Gamma _{03}^{0}=\Gamma _{30}^{0}=\frac{1}{V}\frac{\partial V}{\partial z},
\label{5}
\end{equation}

\begin{equation}
\Gamma _{00}^{3}=c^{2}\frac{V}{U^{2}}\frac{\partial V}{\partial z},
\label{6}
\end{equation}

\noindent and,

\begin{equation}
\Gamma _{33}^{3}=\frac{1}{U}\frac{\partial U}{\partial z}.  \label{7}
\end{equation}

\noindent All other Christoffel symbols are zero.

With these Christoffel symbols we obtain,

\begin{equation}
R_{00}=\frac{V}{U^{3}}\frac{\partial U}{\partial z}\frac{\partial V}{
\partial z}-\frac{V}{U^{2}}\frac{\partial ^{2}V}{\partial z^{2}}  \label{8}
\end{equation}

\noindent and

\begin{equation}
R_{33}=\frac{1}{V}\frac{\partial ^{2}V}{\partial z^{2}}-\frac{1}{UV}\frac{%
\partial U}{\partial z}\frac{\partial V}{\partial z}.  \label{9}
\end{equation}

\noindent All other components of the Ricci tensor are equal to zero. Since
the Ricci tensor is zero in source free space we obtain,

\begin{equation}
\frac{1}{U}\frac{\partial U}{\partial z}-\frac{1}{\frac{\partial V}{\partial
z}}\frac{\partial ^{2}V}{\partial z^{2}}=0  \label{10}
\end{equation}

\noindent for both Eqs. (\ref{8}) and (\ref{9}). The solution to Eq. (\ref
{10}) is\cite{Rohrlich},\cite{Tilbrook}

\begin{equation}
U=\frac{1}{\alpha }\frac{\partial V}{\partial z}  \label{11}
\end{equation}

\noindent where $\alpha $ is a constant, which is easily obtained with
dimensional analysis once $V\left( z\right) $ is chosen. Thus the static
homogeneous field line element becomes

\begin{equation}
c^{2}d\tau ^{2}=V^{2}c^{2}dt^{2}-\frac{1}{\alpha ^{2}}\left( \frac{\partial V%
}{\partial z}\right) ^{2}dz^{2}.  \label{12}
\end{equation}

\noindent In order to calculate the equations of motion one chooses a $%
V\left( z\right) $ that is consistent with this metric. Two examples that
exist in the literature are illustrated in the Lass\cite{Lass} line element

\begin{equation}
c^{2}d\tau ^{2}=e^{\frac{2g}{c^{2}}z}c^{2}dt^{2}-e^{\frac{2g}{c^{2}}z}dz^{2}.
\label{13}
\end{equation}

\noindent and the Rindler\cite{Rindler 66} line element

\begin{equation}
c^{2}d\tau ^{2}=\left( 1+\frac{g}{c^{2}}z\right) ^{2}c^{2}dt^{2}-dz^{2}.
\label{14}
\end{equation}

\noindent These line elements describe the special case in which motion
occurs parallel to the $z$-axis only. Due to the invariance of proper time
any metric consistent with Eq. (\ref{12}) can be used to calculate the
difference in elapsed times for the clock paradox.

In what follows the Rindler metric Eq. (\ref{14}) will be used to calculate
these times. It is convenient to put this in dimensionless form by letting

\begin{equation}
\tau =\frac{c}{g}\tau ^{\prime },\quad t=\frac{c}{g}t^{\prime },\quad z=%
\frac{c^{2}}{g}z^{\prime },  \label{15}
\end{equation}

\noindent where a prime designates the dimensionless quantity. If we let $g$
be the acceleration due to gravity at the surface of the earth and $c$ be
the speed of light, then a time of 1 is $\sim 1$ year and a distance of 1 is 
$\sim 1$ light year. Dropping the primes the Rindler line element becomes

\begin{equation}
c^{2}d\tau ^{2}=\left( 1+z\right) ^{2}c^{2}dt^{2}-dz^{2}.  \label{16}
\end{equation}

Consider the situation in which a test particle begins at rest at the origin
and falls freely. The accelerating twin can be considered to be holding
position at the origin. This twin's clock measures coordinate time. A clock
that falls freely with the test particle can be identified with the stay at
home twin. The equations of motion are obtained from the geodesic equations%
\cite{Weinberg}

\begin{equation}
\frac{d^{2}x^{\nu }}{d\tau ^{2}}+\Gamma _{\mu \sigma }^{\nu }\frac{dx^{\mu }%
}{d\tau }\frac{dx^{\sigma }}{d\tau }=0,\quad x^{0}=t,\quad x^{3}=z.
\label{17}
\end{equation}

\noindent Using the dimensionless Rindler Christoffel symbols and Eq. (\ref
{17}) the equation of motion for coordinate time is

\begin{equation}
\frac{\partial t}{\partial \tau }=\frac{K}{\left( 1+z\right) ^{2}}
\label{18}
\end{equation}

\noindent Where $K$ is a constant of motion obtained from the initial
conditions. Also from Eq. (\ref{17}) the equation of motion for $z$ is

\begin{equation}
\frac{\partial ^{2}z}{\partial \tau ^{2}}+\left( 1+z\right) \left( \frac{
\partial t}{\partial \tau }\right) ^{2}=0.  \label{19}
\end{equation}

\noindent This can be put into a form that is easily solved by substituting
Eq. (\ref{18}) into Eq. (\ref{19}). With this one obtains

\begin{equation}
\frac{\partial ^{2}z}{\partial \tau ^{2}}+\frac{K^{2}}{\left( 1+z\right) ^{3}%
}=0.  \label{20}
\end{equation}

Next we will solve Eq. (\ref{20}) for the first leg of the \textit{gedanken}
experiment. During the first leg the falling twin begins at the origin and
at rest making $K=1$. With this the solution to Eq. (\ref{20}) is

\begin{equation}
z=-1+\sqrt{1-\tau ^{2}}.  \label{21}
\end{equation}

\noindent This result can then be substituted into Eq. (\ref{18}) to obtain

\begin{equation}
t=\tanh ^{-1}\left( \tau \right) .  \label{22}
\end{equation}

It is noteworthy that this solution to the field equations has a vanishing
Riemann tensor as shown by Desloge\cite{Desloge}. This vanishing of the
Riemann tensor would be expected from the symmetry of the Rindler line
element\cite{Weinberg}, which is indistinguishable from a 2-dimensional
problem, and following Weinberg\cite{Weinberg} cannot be associated with a
''true gravitational field''. From physical considerations this is perhaps
not surprising since there can be no physically real source for a
homogeneous gravitational field.

\section{Round Trip in a Static Homogeneous Field}

In order to calculate the full round-trip time the equations of motion for
the second leg must be obtained. The constant of motion for this leg can be
determined by observing that the falling clock comes to rest when it reaches
a maximum distance in the SHF. Figure 1 is a plot of legs one and two for
the outbound trip. Notice that the falling clock comes to rest at a maximum
distance $D$. With this $K=\left( 1+D\right) $ in Eq. (\ref{20}).\bigskip

\begin{figure}[htbp] 
   \centering
   \includegraphics[width=3.5in]{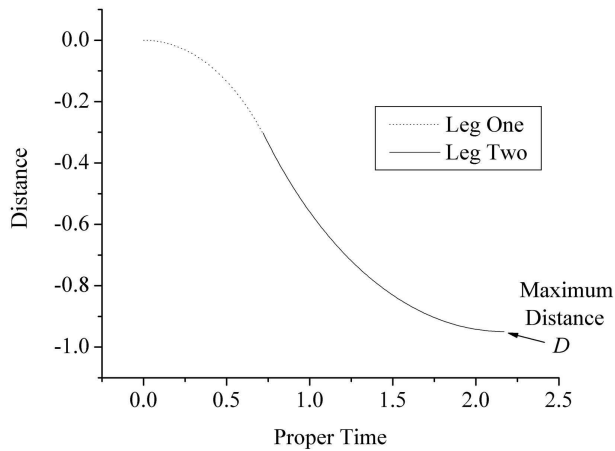} 
   \caption{ This is a plot of the falling twin's distance from the
origin versus proper time for legs one and two of the outbound trip. Notice
that this twin comes to rest at a maximum distance}
   \label{fig:example}
\end{figure}

\bigskip

The equations of motion for leg two can be derived by the same method as the
first leg with the exception that the acceleration has the opposite sign and
the initial conditions must match the position, velocity, proper and
coordinate time at the end of the first leg. The equations obtained in this
way are

\begin{equation}
z=1-\sqrt{\left( 1+D\right) ^{2}-\left( \tau -\sqrt{D\left( D+4\right) }
\right) ^{2}}  \label{23}
\end{equation}

\noindent and

\begin{equation}
t=\tanh ^{-1}\left( \frac{\tau -\sqrt{D\left( D+4\right) }}{1+D}\right)
+2\tanh ^{-1}\left( \frac{\sqrt{D\left( D+4\right) }}{2+D}\right) .
\label{24}
\end{equation}

\noindent Where $D$ is the maximum distance attained by the falling twin. In
order to match the end of the first leg to the beginning of the second leg
the reversal of the direction of acceleration must occur at

\begin{equation}
z=-\frac{D}{2+D}.  \label{25}
\end{equation}

\noindent By letting $z=-D$ in Eq. (\ref{23}) one sees that the total
outbound trip time is

\begin{equation}
\tau _{outbound}=\sqrt{D\left( D+4\right) }  \label{26}
\end{equation}

\noindent which corresponds to a coordinate time of

\begin{equation}
t_{outbound}=2\tanh ^{-1}\left( \frac{\sqrt{D\left( D+4\right) }}{2+D}%
\right) =2\cosh ^{-1}\left( 1+\frac{D}{2}\right) .  \label{27}
\end{equation}

\noindent where the last two expressions in Eq. (\ref{27}) are
mathematically equal due to a hyperbolic functional identity. The return
trip time is equal to the outbound trip time because there is a one-to-one
correspondence between velocity and position for the outbound and inbound
trips, thus the total round-trip time is twice these values.

\section{Special Relativity and Hyperbolic Motion}

The \textit{gedanken} experiment of the clock paradox can also be solved
within the formalism of special relativity\cite{Misner et al}\cite{Rohrlich}%
. Following Misner et al\cite{Misner et al} assign to the accelerating twin
a 4-space velocity $u^{\mu }$ and 4-space acceleration $a^{\mu }=\frac{d}{%
d\tau ^{\prime }}u^{\mu }$ where $u^{\mu }u_{\mu }=-1$ and $a^{\mu }a_{\mu
}=g^{2}$ or in terms of our dimensionless parameters $a^{\mu }a_{\mu }=1$.
The invariance of the square of the 4-space acceleration and the
orthogonality of the 4-space velocity and 4-space acceleration $a^{\mu
}u_{\mu }=0$ leads to a set of differential equations with solution

\begin{equation}
Z=\cosh \left( \tau ^{\prime }\right) -1,  \label{28}
\end{equation}

\noindent and

\begin{equation}
T=\sinh \left( \tau ^{\prime }\right) .  \label{29}
\end{equation}

\noindent Capital letters are used for the coordinates to distinguish these
from the general relativity solution. The proper time in these equations is
the time of a clock moving with the accelerated twin\cite{Misner et al}. The
proper time for the accelerated twin is the same as coordinate time (station
keeping clock) in the SHF Eq. (\ref{27}), by the equivalence principle.
Substituting Eq. (\ref{27}) into Eqs. (\ref{28}) and (\ref{29}) the
equations for the outbound trip distance and time for the stay at home twin
can be written as

\begin{equation}
Z_{outbound}=2\cosh \left( \cosh ^{-1}\left( 1+\frac{D}{2}\right) \right)
-1=D,  \label{30}
\end{equation}

\noindent and

\begin{equation}
T_{outbound}=2\sinh \left( \cosh ^{-1}\left( 1+\frac{D}{2}\right) \right) =%
\sqrt{D\left( D+4\right) }.  \label{31}
\end{equation}

\noindent Comparing Eqs. (\ref{26}) and (\ref{27}) to Eqs.\thinspace (\ref
{30}) and (\ref{31}) demonstrates that the formalisms of special and general
relativity have the same solutions for elapsed times and distances.

\section{Elapsed Time and Distance in a Lorentz Inertial Frame}

The clock paradox \textit{gedanken} experiment has been solved here using
the destination trip distance in the SHF as the parameter. This permits a
simple coordinate transformation from the SHF to a Lorentz inertial frame
(LIF) at the end of leg 2 where the twins are comoving with the destination
star. This transformation can be written as

\begin{equation}
T_{outbound}^{\prime }=\left( 1+D\right) \sinh \left( t_{outbound}\right) ,
\label{32}
\end{equation}

\noindent and

\begin{equation}
L_{outbound}^{\prime }=\left( 1+D\right) \cosh \left( t_{outbound}\right) -1.
\label{33}
\end{equation}

\noindent By equating the square line element in the SHF and LIF 
\begin{equation}
-ds^{2}=dT^{\prime 2}-dx^{2}-dy^{2}-dZ^{\prime 2}=\left( 1+z\right)
dt^{2}-dx^{2}-dy^{2}-dz^{2}.  \label{34}
\end{equation}

\noindent Eqs. (\ref{32}) and (\ref{33}) can be shown to be the correct
transformation from the SHF to the LIF by direct substitution into Eq. (\ref
{34} ).

\begin{figure}[htbp] 
   \centering
   \includegraphics[width=3.5in]{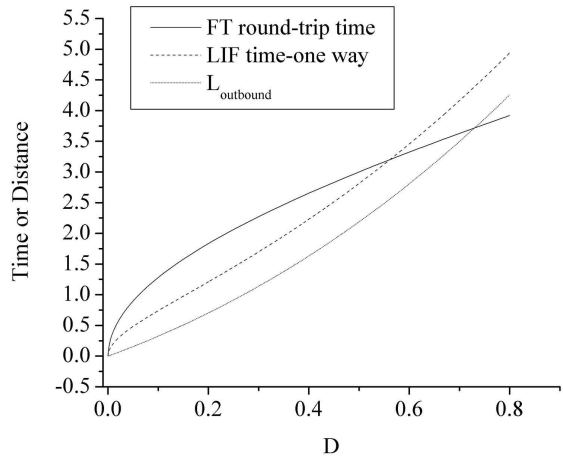} 
   \caption{FT round-trip time is a plot of 2 times Eq. (\ref{26})
and provides the round trip proper time for the freely falling twin. LIF
time one-way is a plot of the outbound trip coordinate time in a Lorentz
inertial frame Eq. (\ref{32}). $L_{outbound}$ is a plot of the distance of
the destination in the LIF Eq. (\ref{33}).}
   \label{fig:example}
\end{figure}

\bigskip

Figure 2 provides a plot of elapsed times and distances in both the SHF and
LIF for distances up to $D=0.8$. In this Figure the outbound coordinate time
in the LIF has been plotted from Eq. (\ref{32}). The round-trip proper time
is plotted in the same Figure as 2 times Eq. (\ref{26}). Examination of this
plot indicates that the outbound coordinate time in the LIF is greater than
the round-trip proper time in the SHF for a one way trip distance greater
than $D=0.56$ and $L_{ouutbound}=2.6$. The magnitude of the minimum distance
for this to occur can also be found analytically by setting 2 times Eq. (\ref
{26}) equal to Eq. (\ref{32})

\[
2\sqrt{D\left( D+4\right) }=\left( 1+D\right) \sinh \left( 2\cosh
^{-1}\left( 1+\frac{D}{2}\right) \right) 
\]

\noindent and solving for $D=\frac{1}{2}\sqrt{17}-\frac{3}{2}$.\bigskip

Figure 3 provides a plot of elapsed times in the SHF and for distances in
the LIF up to $D=16$. In order to fit the one-way distance in the plot $%
L_{outbound}$ has been divided by 100. Examination of this plot indicates
that for distance greater than $L_{outbound}=2,500$ the time differences
between the stay at home and the accelerating twin become
appreciable.\bigskip

\begin{figure}[htbp] 
   \centering
   \includegraphics[width=3.5in]{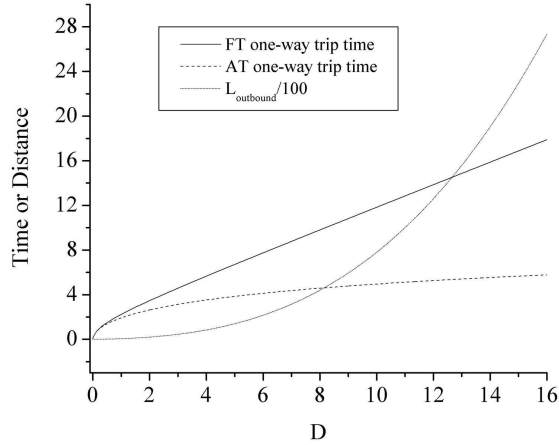} 
   \caption{In order to show the LIF one-way distance the plot shows 
$L_{outbound}$ divided by 100. FT one-way trip time is the one-way (SHF)
falling clock time. AT one-way trip time is the one-way trip (SHF)
coordinate time Eq. (\ref{27}).}
   \label{fig:example}
\end{figure}

\bigskip

The plot of Figure 3 extends to a distance that is $\sim 1/10$ the distance
from the earth to the center of the Milky Way galaxy (assuming an
acceleration of $g=1$). At this distance the time difference between the two
twins is about $20$ years. However, it would seem that the first generation
to achieve this sustained acceleration would have access to $\sim 10^{7}$
star systems in this time interval.

\section{Conclusions}

The \textit{gedanken} experiment of the clock paradox has been solved
exactly, by parameterizing the solution in terms of the maximum trip
distance. The solution was arrived at independently using the formalisms of
special and general relativity and these solutions are shown to be
identical. We have also shown that for a one-way trip of sufficient distance
the outbound trip coordinate time in the LIF is greater than the round-trip
proper time in the SHF. Transforming the maximum trip distance and one-way
elapsed time in the SHF to a LIF the elapsed times and distances for
realistic journeys to the stars were calculated.

\end{document}